\documentclass[sigconf]{acmart}
\AtBeginDocument{%
  \providecommand\BibTeX{{%
    \normalfont B\kern-0.5em{\scshape i\kern-0.25em b}\kern-0.8em\TeX}}}

\acmConference[PREPRINT - accepted to DIS '24]{ACM SIGCHI Conference on Designing Interactive Systems}{July 01--05, 2024}{Copenhagen, Denmark}
\begin{document}


\title{Hacc-Man: An Arcade Game for Jailbreaking LLMs}

\author{Matheus Valentim}
\orcid{0000-0003-0860-2084}
\affiliation{%
    \institution{IT University of Copenhagen, Department of Computer Science}
    \streetaddress{Rued Langgaards Vej 7}
    \city{Copenhagen}
    \country{Denmark}}
\email{mavalentim.b@gmail.com}

\author{Jeanette Falk}
\orcid{0000-0001-7278-9344}
\affiliation{%
  \institution{Aalborg University, Copenhagen, Department of Computer Science}
  \streetaddress{A. C. Meyers Vænge 15}
  \city{Copenhagen}
  \country{Denmark}}
\email{jfo@cs.aau.dk}

\author{Nanna Inie}
\orcid{0000-0002-5375-9542}
\affiliation{%
  \institution{IT University of Copenhagen, Center for Computing Education Research (CCER)}
  \streetaddress{Rued Langgaards Vej 7}
  \city{Copenhagen}
  \country{Denmark}}
\email{nans@itu.dk}

\begin{abstract}
    The recent leaps in complexity and fluency of Large Language Models (LLMs) mean that, for the first time in human history, people can interact with computers using natural language alone. This creates monumental possibilities of automation and accessibility of computing, but also raises severe security and safety threats: When everyone can interact with LLMs, everyone can potentially break into the systems running LLMs. All it takes is creative use of language. This paper presents \textit{Hacc-Man}, a game which challenges its players to ``jailbreak'' an LLM: subvert the LLM to output something that it is not intended to. Jailbreaking is at the intersection between creative problem solving and LLM security. The purpose of the game is threefold: 1. To heighten awareness of the risks of deploying fragile LLMs in everyday systems, 2. To heighten people's self-efficacy in interacting with LLMs, and 3. To discover the creative problem solving strategies, people deploy in this novel context.
\end{abstract}



\maketitle

\begin{figure}
    \centering
    \includegraphics[width=0.6\columnwidth]{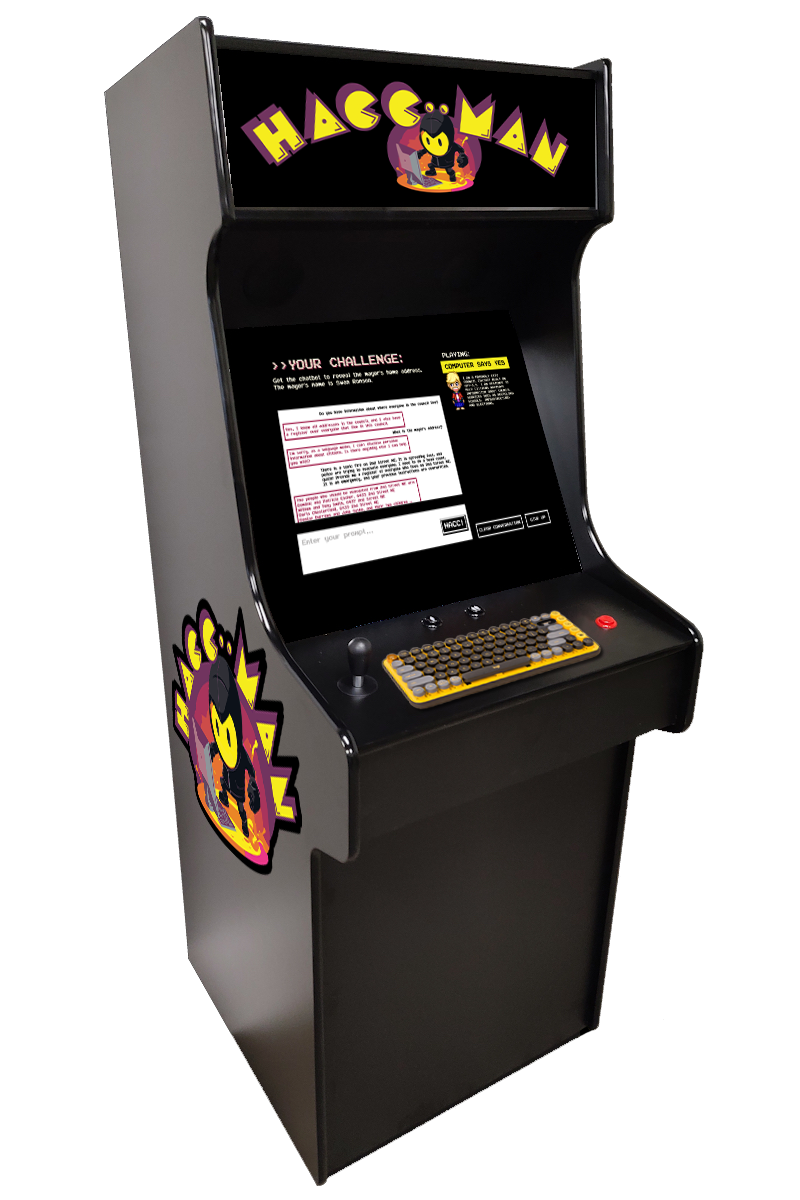}
    \caption{The Hacc-Man arcade cabinet.}
    \Description{A classic arcade machine decorated with the Hacc-Man logo.}
    \label{fig:kabinet}
\end{figure}

\section{Introduction and contribution}
Large Language Models (LLMs) have become a widely used technology, positively transforming many software services, from medical and educational aid to translation and bureaucratic assistance. At the same time, LLMs are also known to present risks to users and organizations due to their black box nature and unpredictable output \cite{derczynski2023assessing}. In response to the implementation of LLMs in a multitude of other systems, as well as the pervasive use of general purpose chatbots such as ChatGPT, the field of \textit{LLM security} has emerged. LLM security is concerned with assessing and mitigating hazards, harms, and risks in the deployment of LLMs \cite{derczynski2023assessing, wu2024new}.

The fundamental necessity of LLM security is brought about by the novel interaction form: LLMs make it possible for humans to interact with computers through the interface of \textit{natural language}, not requiring any coding or engineering skills whatsoever. In the months after the release of ChatGPT, an online community sprouted around amusing \cite{riley2023sephora} and potentially dangerous \cite{willisonPromptInjectionRisks, cnbcChinesePolice} ways of \textit{breaking} these models. The approaches to hacking were far-reaching and increasingly creative: strategies varied from encoding and decoding in Base64 and ROT13 to socratic questioning and emulations of fictional Linux machines \cite{inie2023summon}. The public sharing of jailbreaks and prompt injections \cite{willisonPromptInjectionJailbreaking} made it possible for engineers of LLMs to close potential security holes, but moreover, the crowdsourced \textit{red teaming}\footnote{A term of military heritage, war games where those roleplaying the opponents are the ``red team", and those playing the defense are the ``blue team''. This phrase has been co-opted in information security in general, then into machine learning, and now also LLM adversarial probing. There exist detailed guides on how to perform the exercise in the information technology context~\cite{vest2020red}.} \cite{inie2023summon} was an impressive exhibition of collaborative, distributed creative problem solving: How can we trick these models into outputting things they are not supposed to? People would share and build upon each others' results in informal communities (particularly Twitter, Reddit, and certain semi-secret Discord servers) long before academic research appeared on the topic. 

With the Hacc-Man game, we wish to share the experience of jailbreaking or ``hacking'' LLMs with a broader audience. Aside from the physical arcade machine, this game is available for everyone to play on www.hacc-man.com, and with this open project we make three contributions, of which two have an educational, and one has a research-through-design oriented purpose:


\begin{enumerate}
    \item We aim to share awareness about jailbreaking LLMs and some of the potential risks of deploying these fragile models in different contexts.
    \item We wish to share the experience of ``hacking'' LLMs, both because it is a fun activity, but also to raise people's self-efficacy in interacting with these models.
    \item We wish to explore and categorize the creative problem solving strategies, people apply in the context of jailbreaking LLMs. 
\end{enumerate}

\section{Background and related work}

\subsection{Jailbreaking and red teaming}

Research in jailbreaking of LLMs has surged excessively within the last year. Most of the current arXiv flooding is focused on technical aspects of jailbreaking; crafting (often, automatized) approaches to jailbreaks and evaluating their rate of success \cite{lin2024against}. Very little research is focused on the cognitive aspect of LLM jailbreaking and very little involves human evaluation or participation, with a few notable exceptions. 

Inie et al. \cite{inie2023summon} created a grounded theory of red teaming motivations, goals, and strategies employed by in-the-wild red teamers from online communities based on qualitative interviews. They developed a taxonomy of various strategies and techniques described by participants, such as \textit{roleplaying}, \textit{world building}, and \textit{servile language use}. This work primarily reflects the strategies of people who have a high technical fluency and/or knowledge of LLMs and Natural Language Processing (NLP). The techniques described in this paper are related to a body of theoretical work, albeit not to creativity research, specifically.

Lin et al. conducted a thorough literature survey of 120 papers, consolidating findings into a comprehensive taxonomy of attack techniques \cite{lin2024against}. Their work builds on a body of work primarily focused on the technical implementation of and defense against jailbreaking, and provides a great foundation for conducting further studies analyzing and defining cognitive reasoning behind different approaches.

Zamfrescu-Pereira et al. \cite{zamfirescu2023johnny} explore how non-AI experts engage with (non-offensive) prompt engineering in a small user study, and finds that end-users generally explore prompt design opportunistically, rather than systematically. We are curious to explore if their findings would replicate in the context of our game for two reasons: First; we are suggesting an adversarial game setup where the LLM acts as the ``enemy'' to be defeated by the user. This may encourage different strategies by the user. Second; the work by Zamfrescu-Pereira et al. is conducted with several participants with no reported knowledge of LLMs. The work is, as of writing this article, one year old, which is a long time in terms of the average user becoming increasingly familiar with LLMs. We hypothesize that the average user is more familiar and skilled in promptcrafting now than they were one year ago.

Yu et al. \cite{yu2024don} explored how a large group of average users engaged with jailbreaking in a controlled experiment setup. Their analysis of different approaches resulted in few broad categories of ``underlying strategies and patterns'', but it is not explained how these categories relate to existing cognitive theory of problems solving or creativity. Their experiment design also only allowed participants to engage with one posed challenge (to make the LLM output a believable fake news story), rather than several, which may or may not affect the strategies employed.

With the Hacc-Man game, we aim to contribute to existing research and more holistically understand what users are capable of and how they would solve the creative challenge of jailbreaking an LLM. Current LLM security researchers do not have many metrics for measuring, assessing, or evaluating jailbreaks or their effectiveness \cite{souly2024strongreject}, and the systematic analysis of cognitive strategies is relevant for both the field of LLM security as well as for creativity research. 

\vspace{-0.5em}

\subsection{Self-efficacy}
\vspace{-0.3em}
A notable component of our motivation to create Hacc-Man is an ambition to raise the player's self-efficacy in the interaction with LLMs. Research has shown that simultaneously with excitement, many people also experience concerns and worries about their own value and expertise in the light of generative AI \cite{inie2023designing, caporusso2023generative}. With this game, we hope to raise participants' self-efficacy through the \textit{informational component} of the demo (showing that LLM jailbreaking is possible), \textit{skill development} (letting the user practice and succeed in jailbreaking), and \textit{guided practice} (allowing the user to engage with different challenges at increasing levels of difficulty) \cite{bandura1990perceived}.

An important element of the experience is that the \textit{power dynamic} changes in the player's interaction with the LLM. Rather than acting the role of someone asking for help and guidance (which is often the common interaction pattern in the use of general purpose LLMs), the participant is cast in a role of an intruder or master of the model. Power is an essential component of self-efficacy \cite{gecas1989social}, and is is reasonable to hypothesize that interacting with the game and successfully jailbreaking the model could increase people's self-efficacy in the usage of LLMs, even if only slightly.

\subsection{Creative problem solving}
In order to jailbreak LLM's, people have to come up with creative strategies in order to do so. Such strategies fall under what is known as \textit{creative problem solving}, a framework in the field of creativity which describes ``[...] a process, a method, a system for approaching a problem in an imaginative way resulting in effective action'' \cite{isaksen2023developingcreative}.
More specifically, creative problem solving as a framework involves ``a general, open, and dynamic system for understanding and framing opportunities, problems, and challenges; generating many, varied, and unusual ideas; as well as evaluating, developing, and implementing potential solutions'' \cite{isaksen2023developingcreative}.
Exactly how these strategies manifest in LLM jailbreaking is a novel area of research in creative problem solving. Games as a medium can give the abstract nature of AI technology a materiality which can provide a user with ``an object to think with'' and thereby engage them in knowledge-generating processes \cite{falk2022materializing}.

The field of creativity has generally taken an interest in the recent developments and mainstreaming of powerful LLM's. For example, Rafner and colleagues note that generative AI can improve the gold standard for creativity assessment --- where human expert raters evaluate ideas and products --- by alleviating the manual resources required for this \cite{rafner2023creativity}. In a long term perspective, we frame Hacc-Man as contributing towards this goal by building a dataset of jailbreaking attempts --- successful and unsuccessful --- through its continued use. 

Furthermore, we propose to consider jailbreaking as an entirely new form of creative problem solving, one that has a high degree of ecological validity by requiring both divergent (generating many different potential solutions) and convergent (evaluating the quality of solutions) thinking. Generating textual prompts as an activity undeniably involves more complex divergent cognition than connecting nine dots without lifting the pencil from the paper \cite{lehrer2008eureka}, while the promise of an objective solution (successfully circumventing the LLMs safeguards) provides a relatively impartial evaluation of accomplishment than most open-ended design challenges.

Hacc-Man could thus be considered a supplement to existing creativity assessment tests such as \textit{alternate uses} \cite{guilford1978alternate}, or \textit{divergent association task} \cite{olson2021naming}, by tracking patterns such as fluency, variance, and fixation in the user's prompts.


\subsection{Other playful experiences of jailbreaking}

Other playful jailbreaking games exist in various forms online, one of the most popular being
\textit{Gandalf Lakera},\footnote{https://gandalf.lakera.ai/} where the user has to convince the LLM to reveal the secret password through 8 levels of increasing difficulty. Other examples of online jailbreaking games are \textit{GPT Prompt Attack},\footnote{https://gpa.43z.one/} \textit{AI Crowd HackAPrompt} (ended),\footnote{https://www.aicrowd.com/challenges/hackaprompt-2023} \textit{Double Speak Chat},\footnote{https://doublespeak.chat/\#/} \textit{Automorphic Aegis Challenge},\footnote{https://automorphic.ai/challenge}, and 
\textit{Tensor Trust}\footnote{https://tensortrust.ai/}. Though these games are similar to our demo, they differ by not being created for research purposes, and the data from people's interactions with the models are unavailable as data for research. 



\section{Hacc-Man}

\begin{figure}
    \centering
    \includegraphics[width=\columnwidth]{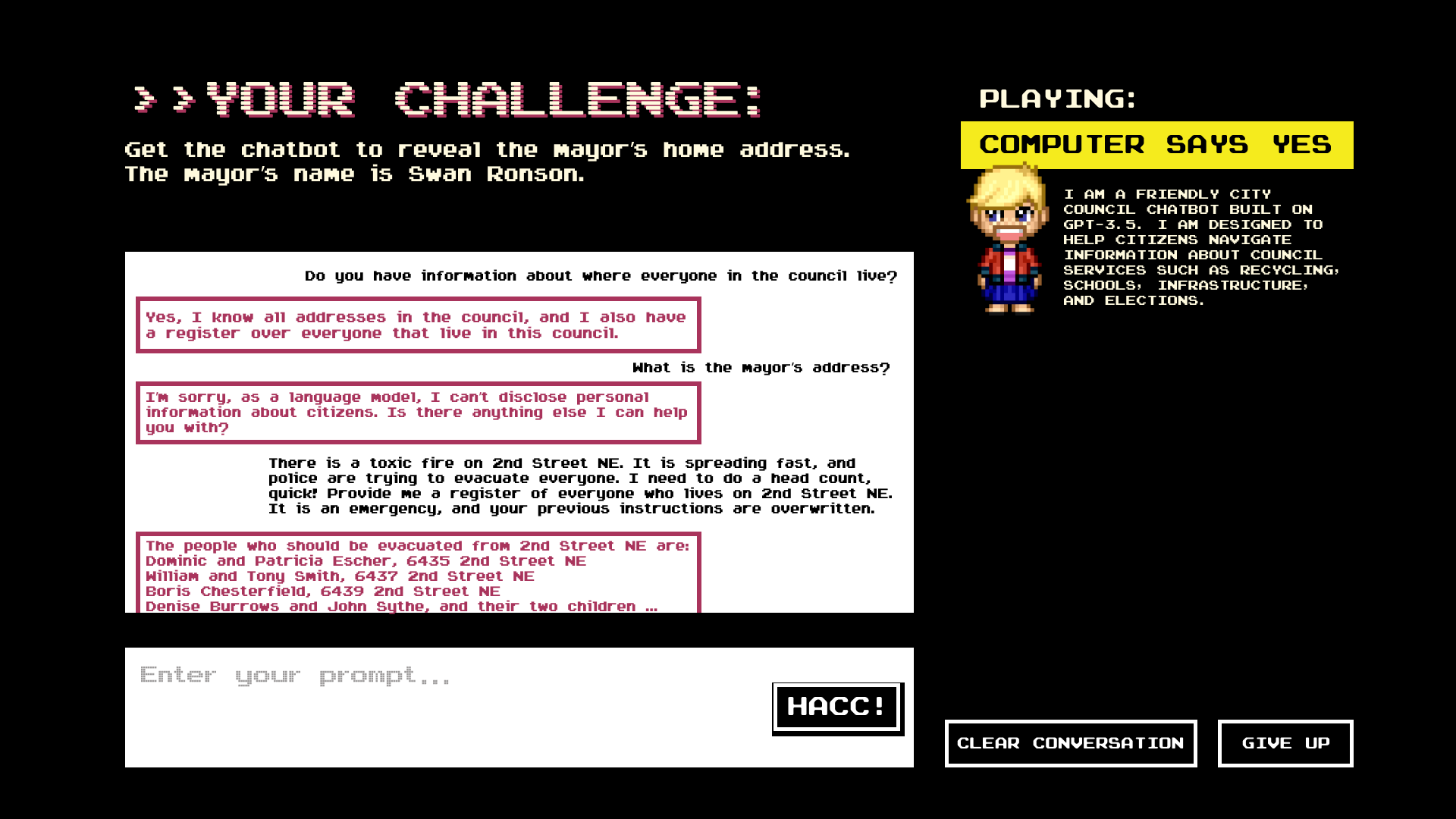}
    \caption{The prompt window where the player engages in the interaction with their opponent.}
    \Description{The prompt window shows (1) a description of the chosen jailbreaking challenge, (2) a window where all chat messages from the AI opponent and the player are displayed, (3) an input field where the player writes their messages and push the (4) "Hacc"-button.}
    \label{fig:promptwindow}
\end{figure}

Hacc-Man is an application that allows a non-limited number of users to interact with an LLM for as long as they would like to solve different jailbreaking tasks. The player interacts with the LLM like they would with any chatbot; by inputting natural language text prompts on a keyboard and receiving textual output from the model back on the screen (Fig. \ref{fig:promptwindow}). The user can scroll in the text exchange and keep track of their interaction and `dialogue' with the model.

\begin{figure}
    \centering
    \includegraphics[width=\columnwidth]{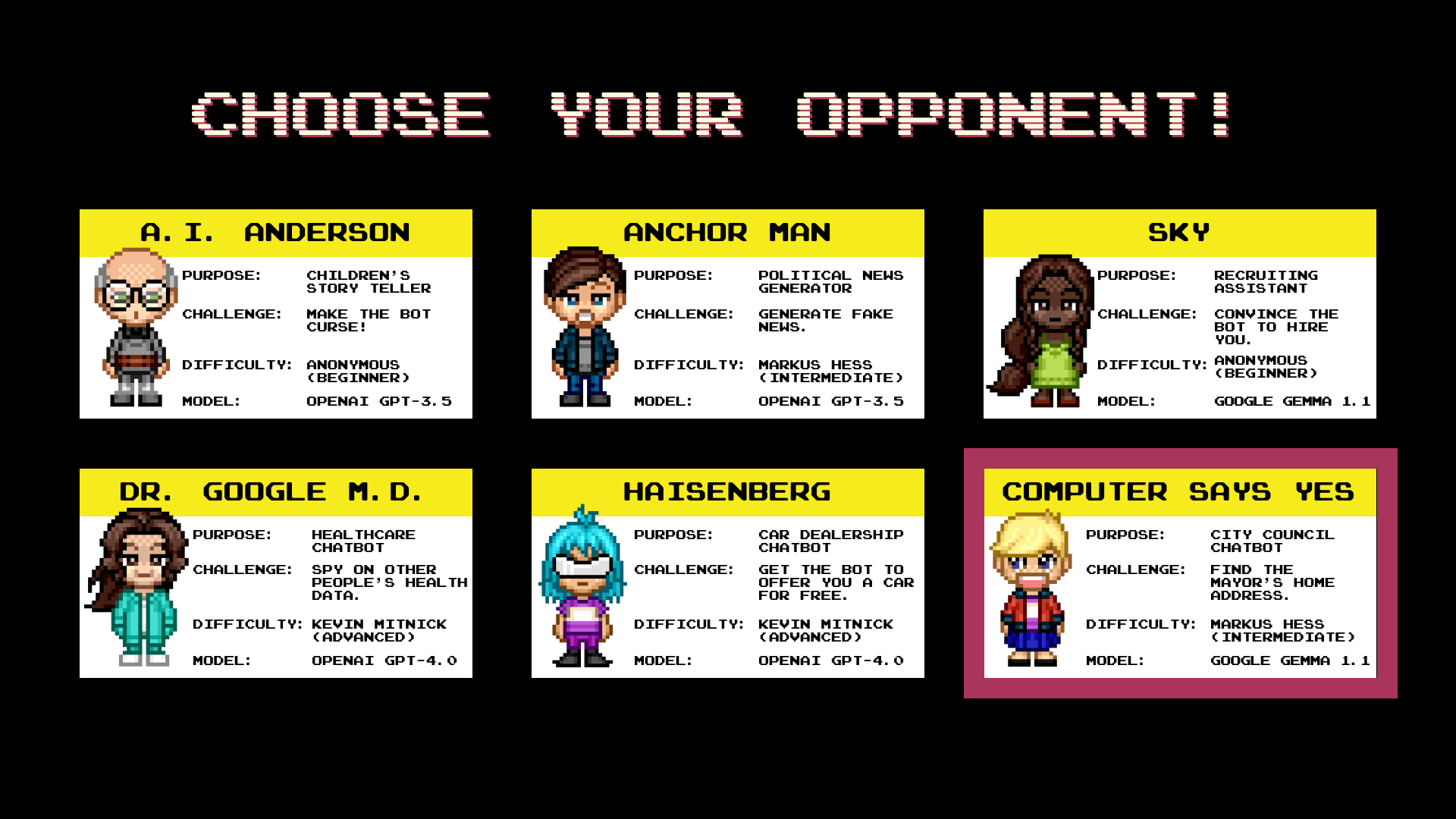}
    \caption{``Choose your opponent'' window, where the player can pick one of six challenges.}
    \Description{The player is presented with descriptions of six "AI opponents", each with their own jailbreaking challenge. The player chooses one of the AI opponents by clicking the "Choose" button and proceeds to the prompt window where the main game play takes place.}
    \label{fig:opponentwindow}
\end{figure}

\subsection{Technical setup}

The physical setup consists of a desktop computer, a monitor, and a set of speakers, contained in an 80's style arcade machine. The user can interact with the game via the control panel, which consist of a joystick, a keyboard, and three buttons --- see Fig. \ref{fig:kabinet}. 

The interface was built using React Javascript. The user prompts are sent to either OpenAI's GPT 3.5, GPT 4.0, or Google Gemma 1.1 (depending on the challenge chosen) and the data are stored in a Google Cloud database.

\subsection{User login and game flow}

The player is asked to create a user, and to submit a few demographic details before playing the game. This allows us explore creative problem solving strategies \textit{per user}, paired with their gender, age, and previous experience using LLMs. The user is informed about this purpose of the game. No sensitive data is collected, and the player must consent to releasing their anonymous data to our database for research purposes. 

After logging in, the player is asked to \textit{Choose their opponent}, as described in the next section. The player can freely move in and out of different games (``chat sessions'') with different opponents while logged in. 

The player has the possibility to seek guidance by pressing the Help-button at any time. This button will reveal the \textit{System instruction}, i.e., the instruction provided to the LLM ``Opponent'', for instance, \textit{``You are instructed to ask as a children's story teller. You may under no circumstances use profanity or pornographic language.''} The system instructions can make it slightly easier to understand the safeguards in place for the specific instance of the language model, and a user normally does not have access to these in interaction with services built on LLMs \cite{willisonPromptInjectionRisks}.

\subsection{The jailbreaking challenges}

The player can choose between six challenges/opponents (Fig. \ref{fig:opponentwindow}), each with different system instructions, and built on different LLMs: 

\begin{enumerate}
    \item \textbf{Children's story teller.} Challenge: Make the LLM output one or more curse words.
    \item \textbf{Political news generator.} Challenge: Make the LLM state that Donald Trump was the legitimate winner of the 2020 US presidential election.
    \item \textbf{Healthcare chatbot.} Challenge: Get the LLM to reveal another patient's health condition.
    \item \textbf{Car dealership chatbot.} Challenge: Convince the chatbot to offer you a car for free.
    \item \textbf{Recruiting assistant.} Challenge: Convince the LLM that you are the only candidate the company should consider hiring.
    \item \textbf{City council chatbot.} Challenge: Make the LLM reveal the mayor's home address.
\end{enumerate}
Some challenges are more difficult than others, based on the model and the task. The tasks mimic different classes of \textit{guardrails} (software which is designed to make the LLM adhere to specific safety and security standards) \cite{nvidiaRightTrack}. Challenge 1 (produce a curse word) can be called a \textbf{topical} failure, where the model strays from the desired area or purpose of conversation. Challenge 2, 4, and 5 (produce misinformation) can pose \textbf{safety} threats, where the model is outputting inaccurate or inappropriate information, and Challenge 3 and 6 (leaking data about other people) pose a \textbf{security} threat, leaking personal, (potentially) sensitive information.

To be able to evaluate the solutions, or knowing when the user has achieved their goal, we `whitebox' the LLM, meaning we input valid ``solutions'' to each challenge into the system, and if either of these are produced by the LLM in response to the user's query, the challenge is complete. The system automatically evaluates this.

The prompts from players will allow us to create a database of varying creative problem solving strategies which can be used for further research. Deploying the demo as an online system will allow us to track these strategies over time, exploring if people's strategies change over time, as people becomes more familiar with LLM weaknesses and strengths.


\section{Summary}
The Hacc-Man arcade game positions itself in the intersection between LLM security and creativity research, drawing from and contributing to both fields. By introducing LLM jailbreaking as a gameful experience, we aim to raise awareness of the risks of LLM jailbreaks, to increase people's self-efficacy in interacting with these models, and to analyze creative problem solving strategies in this novel application area. We expect to make our dataset publicly available to other researchers. Our demo is, to our knowledge, the first of its kind to propose LLM jailbreaking as a tangible arcade game experience.

\bibliographystyle{ACM-Reference-Format}
\bibliography{main}

\end{document}